\documentclass[prl,showpacs,twocolumn,superscriptaddress]{revtex4}
\usepackage{epsfig}
\usepackage{amsmath}
\usepackage{graphicx}   % Include figure files
\usepackage{dcolumn}    % Align table columns on decimal point
\usepackage{longtable}
\usepackage{bm}         % bold math
\usepackage{color}
\usepackage{multirow}

\usepackage{gensymb}

\begin{document}

\draft
\title{Atomic diffusion in $\alpha$-iron across the Curie point: an efficient and transferable ab-initio-based modelling approach}

\author{Anton Schneider}
\affiliation{DEN-Service de Recherches de M\'etallurgie Physique, CEA, Universit\'e Paris-Saclay, F-91191 Gif-sur-Yvette, France}

\author{Chu-Chun Fu}
\affiliation{DEN-Service de Recherches de M\'etallurgie Physique, CEA, Universit\'e Paris-Saclay, F-91191 Gif-sur-Yvette, France}

\author{Fr\'ed\'eric Soisson}
\affiliation{DEN-Service de Recherches de M\'etallurgie Physique, CEA, Universit\'e Paris-Saclay, F-91191 Gif-sur-Yvette, France}

\author{Cyrille Barreteau}
\affiliation{DRF-Service de Physique de l'Etat Condens\'e, CEA-CNRS, Universit\'e Paris-Saclay, F-91191 Gif-sur-Yvette, France}

\date{\today}

\begin{abstract}
An accurate prediction of atomic diffusion in Fe alloys is challenging due to thermal magnetic excitations and magnetic transitions. We propose an efficient approach to address these properties via Monte Carlo simulation, using \textit{ab-initio} based effective interaction models. The temperature evolution of self- and Cu diffusion coefficients in $\alpha$-iron are successfully predicted, particularly the diffusion acceleration around the Curie point, which requires a quantum treatment of spins. We point out a dominance of magnetic disorder over chemical effects on diffusion in the very dilute systems.
\end{abstract}

\maketitle

%(main body = main text + formula + figures + tables < 3750 words, typically 4 figures and around 4 pages. References and acknowledgement are not counted)

%Context and motivation :

Atomic diffusion plays a central role dictating the kinetics of numerous physical processes in solids, such as surface and interfacial segregation, precipitation and phase transitions. Iron based alloys, being the basis of steels, are certainly one of the most studied systems from both theoretical and experimental points of view. Experimental data for solute and solvent diffusion in iron alloys are usually known at rather high temperatures only (above 750K)\cite{Hettich1977, Buffington1961, Walter1969, Iijima1988, James1966, Borg1960, Geise1987, Graham1963, Kucera1974, Luebbehusen1988, Speich1966, Rothman1968, Lazarev1970, Salje1977}, whereas a quantitative modeling of the diffusion coefficients as functions of temperature in these systems is not obvious, since the effects of thermal magnetic excitations and magnetic order-disorder transitions need to be properly described.

Even the simplest case of self-diffusion in body-centered-cubic (bcc) iron via the vacancy mechanism is still the focus of various recent modelling efforts \cite{Huang2010, Ding2014, Sandberg2015, Wen2016, Tapasa2007}. To the best of our knowledge, there were very few theoretical studies on solute diffusion in iron across the Curie point \citep{Huang2010, Ding2014, Wen2016}. To address these properties at a thermal-vacancy regime, vacancy formation and migration free energies should be determined, as the diffusion of Fe and most substitutional solutes in iron is ruled by first nearest neighbor ($1nn$) atom-vacancy exchanges.

Density functional theory (DFT) calculations provide an accurate estimation of the vacancy properties in the ground-state ferromagnetic (FM) bcc iron \citep{Sandberg2015, Ding2014, Huang2010, Domain2002, Olsson2007}. But, theoretical determination of these properties at increased temperatures, when magnetic excitations emerge, becomes non-trivial \citep{Ruban2012,Gambino2018,Gorbatov2011,Gorbatov2016}.
So far, many previous studies have estimated the temperature evolution of the diffusion activation energy $Q(T)$ using the Ruch model \citep{Ruch1976}, with that, the diffusion coefficients $D(T)=D_0exp(\frac{-Q(T)}{k_{B}T})$ can be obtained.

At a given temperature, $Q(T)$ is actually the magnetic free energy of activation, which includes the contributions from both the vacancy formation and the various atom-vacancy exchange barriers. In the case of self-diffusion, it can be written as a sum of the vacancy formation and the vacancy-atom exchange magnetic free energy ($Q(T) = G^{f}_{mag} + G^{m}_{mag} $). These values are called "magnetic" free energies because only the magnetic entropy is included.

The Ruch model \citep{Ruch1976} proposes:

\begin{equation}
\label{ruch_eq}
Q(T) = Q_{PM} (1 + \alpha S^{2})
\end{equation}

where $Q_{PM}$ is the activation enthalpy of a perfect paramagnetic (PM) state, where no more magnetic correlation is present. $S$ is the magnetic order parameter (the reduced magnetization) and $\alpha$ is a scalar. Note that to apply the Ruch model, the temperature dependence of $S$, and either $Q_{PM}$ or the $\alpha$ parameter should be known, in addition to $Q_{FM}$ (the activation enthalpy of the ideal FM state). An example of such application of the Ruch model can be found in Ref. \citep{Versteylen2017}.
 
For a perfect PM state, the vacancy formation and migration enthalpies, and therefore $Q_{PM}$ were often estimated via DFT, by adopting for instance the disordered local moment (DLM) approach within a collinear approximation \citep{Sandberg2015,Gambino2018}, or via an expansion on a set of spin spirals \citep{Ding2014,Ruban2012}. In any case, the magnetic short-range order (MSRO) was generally not considered. Such PM states are therefore expected only at extremely high temperatures.
The temperature dependence $S$ for a pure or a dilute system can be easily provided by experiments or simple Ising or Heisenberg models. It is however more difficult to obtain for concentrated alloys with any microstructure.

Concerning intrinsic approximations of the Ruch model\citep{Ruch1976}, it is derived from the Ising model, and due to a mean field approximation, the MSRO effect is absent. Therefore, $Q(T) = Q_{PM}$ immediately above the Curie point.
 
Another approach used to determine the temperature evolution of the diffusion  properties is the spin-lattice dynamics \citep{Wen2013, Wen2016}, employing empirical potentials and Heisenberg-interaction terms. In particular, a recent study of Wen et al. \citep{Wen2016} reported a detailed investigation of self-diffusion in bcc iron. A major advantage of such an approach consists in the natural inclusion of the combined phonon-magnon effects. However, in practice an accurate potential is not obvious to parameterize, especially for magnetic alloys with structural defects. Furthermore, such spin-lattice dynamics simulations can hardly reach very long time scales and large simulation systems (typically a few tens of nanoseconds and 16000 atoms as in Ref. \onlinecite{Wen2016}).

The present study aims at proposing an efficient and quantitative modelling approach that enables a continuous prediction of diffusion properties versus temperature, including explicitly spin and atomic variables. 
A DFT-based effective interaction model (EIM) coupled with on-lattice Monte Carlo (MC) simulations is adopted. We consider the case of self- and Cu diffusion in bcc iron to illustrate the ability of the methodology to predict diffusion properties, which can be transferred to other magnetic metal alloys. 

Some previous studies have already proposed EIMs with both magnetic and chemical variables \citep{PierronBohnes1983,Lavrentiev2010,Gorbatov2013,Lavrentiev2014,Lavrentiev2016,Ruban2012,Chapman2019}, but without considering defects and, therefore, not able to study diffusion. On the other hand, more conventional EIMs have been developped to study thermodynamic and kinetic properties without explicitly including spin variables \cite{Soisson2007, Martinez2012, Senninger2014, Levesque2011}.   
The present EIM (Eq. \ref{EIM}) consists in a magnetic part with a Landau-Heisenberg form as in Ref. \citep{Lavrentiev2010,Lavrentiev2014,Lavrentiev2016,Ruban2007}. This allows to account for both longitudinal and transversal excitations of spins. In addition, the pairwise terms ($V_{ij}$) capture chemical interactions between atoms.  

%\begin{widetext}

\begin{equation}
\begin{split}
\label{EIM}
H = \sum\limits_{i}^{N} (A_{i}M_{i}^{2} + B_{i}M_{i}^{4}) + \sum\limits_{i}^{N}\sum\limits_{n}^{P}\sum\limits_{j}^{Z_n} J_{ij}^{(n)}{\bf M}_{i}\cdot {\bf M}_{j} \\ + \sum\limits_{i}^{N}\sum\limits_{n}^{P}\sum\limits_{j}^{Z} V_{ij}^{(n)}
\end{split}
\end{equation}

%\end{widetext}

where $Z_n$ is the coordination of the $n$-th neighbor shell and $M_{i}$ is the magnetic moment of the i$th$ atom. $A_{i}$ and $B_{i}$ are the magnetic on-site coefficients.
$J_{ij}^{(n)}$ and $V_{ij}^{(n)}$ denote respectively the magnetic exchange-coupling and the chemical-interaction parameters for $i$ and $j$ atoms being $n$-th nearest neighbors.

First, an EIM for pure iron in a bcc lattice is parameterized on DFT \citep{Bloechl1994,Kresse1999,Kresse1993,Kresse1996a,Kresse1996b,Perdew1999,Chen1995,Monkhorst1976,Methfessel1989,Chen1985, Wolverton2004,Togo2015} data. Fitting the magnetic parameters using DFT consists in evaluating the energy difference between systems with similar atomic configurations but distinct magnetic configurations (See \cite{SM} for details). We checked that the Curie temperature is correctly reproduced ($T_C$ = 1050 K, the experimental value being 1044 K \citep{Keffer1966}).
Then, to include the presence of a vacancy (EIM$^{V}$), the on-site $A$ and $B$ parameters are modified for atoms located at the first and second nearest-neighbor ($1nn$ and $2nn$) sites of the vacancy, in order to reproduce the change of their magnetic moment magnitude, while, for simplicity, the $J_{ij}$ remain unchanged. Vacancy formation energies for distinct magnetic spin configurations around the vacancy predicted by DFT are successfully captured by this simple model \cite{SM}. In order to simulate the atomic migration, another pure-iron derived EIM$^{SP}$ is also constructed to describe the energetics of an Fe atom at a saddle-point position. In this case, the on-site and $J_{ij}$ parameters of the saddle point atom and their $1nn$ and $2nn$ atoms are modified based on DFT data. 
The atom-vacancy exchange barriers are then determined by the energy calculated using the EIM$^V$ and the EIM$^{SP}$. 
Note that such way of barrier determination was intensively applied in previous studies using non-magnetic interaction models \citep{Soisson2007, Martinez2012, Senninger2014}.  

For Cu-diffusion in bcc iron, a Cu atom and a vacancy should be included in the iron system. Similarly, an EIM with all the atoms at lattice positions and another one with an atom (Fe or Cu) at a saddle-point site are parameterized on DFT data on Cu-vacancy binding energies and atom-vacancy exchange barriers with various spin configurations. The numerical parameters of the various EIMs are given in Ref. \cite{SM}. 

The lattice vibrational effects (vibrational entropies and attempt frequencies) are not intrinsically accounted in the present EIM-Monte Carlo set-up but calculated separately by DFT \citep{SM}. The magnon-phonon effects \citep{Koermann2014} are therefore not considered.

The tracer diffusion coefficients can be expressed with the Einstein's formula\citep{Einstein1905,Einstein1906,Smoluchowski1906} with
$<r^{2}>$ and $t$ being the mean square displacement of the tracers 
and the corresponding physical time:

\begin{equation}
\label{selfdiff2}
D^{*} = \frac{<r^{2}>}{6t}
\end{equation}

For the self-diffusion case, it can also be written in terms of the vacancy concentration and the migration barrier at a given $T$ \citep{Adda1966, Mehrer2007} as:

\begin{equation}
\label{selfdiff1}
D^{Fe*}_{Fe} = a^{2}f_{0}C_{v}\nu_{0}\exp(\frac{-G^{m}_{mag}}{k_{B}T})
\end{equation}

where $a$ is the lattice constant, $f_{0}$ is the self-diffusion correlation factor (0.727 for a bcc lattice \citep{Leclaire1970}), $C_{v}$ is the equilibrium vacancy concentration, $\nu_{0}$ is the attempt frequency, $G^{m}_{mag}$ is the magnetic free energy barrier for the vacancy-Fe exchange (vacancy migration), and $k_{B}$ and $T$ are respectively the Boltzmann factor and the absolute temperature. Here $C_{v} = exp(\frac{-G^{f}}{k_{B}T})$, with $G^{f}$ being the vacancy formation free energy. Both magnetic and vibrational entropies are considered in this study, and the latter is calculated via DFT at the FM state. 

Similarly, the solute (Cu) tracer diffusion coefficient in Fe at the dilute limit can be written \citep{Adda1966, Mehrer2007} as: 

\begin{equation}
\label{solute_diff}
D^{Cu*}_{Fe} = a^{2}C^{1nn}_{v}f_{2}\nu_{2}\exp(\frac{-G^{m,Cu}_{mag}}{k_{B}T})
\end{equation}

where $C^{1nn}_{v}$ is the equilibrium vacancy concentration at a $1nn$ site of the solute, $\nu_{2}$ is the vacancy-Cu exchange attempt frequency, $f_{2}$ is the solute diffusion correlation factor and $G^{m,Cu}_{mag}$ is the magnetic free energy barrier for the vacancy-solute exchange.

We propose a Monte Carlo method which allows to determine the vacancy formation magnetic free energy as a function of temperature. Two separate Fe subsystems are considered with the first one frozen at the FM state, while the magnetic configuration of the second one is allowed to evolve according to temperature. The vacancy is allowed to visit each site of the two systems via the Metropolis algorithm. Then, based on the relative number of “visits” to the two systems and the vacancy formation energy at the FM state, which is known, the magnetic free energy of vacancy formation versus temperature can be obtained (More details in Ref. \onlinecite{SM}). Note that, as mentioned in Ref. \citep{Bergqvist2018,Woo2015,Wen2016}, a quantum treatment of spins is necessary for a correct prediction of the magnetic entropy, at low temperatures. We therefore adopted the Bose-Einstein statistics in our spin-MC simulations \citep{SM} up to the Curie point, following the quasi-harmonic approach of Ref. \onlinecite{Bergqvist2018,Wu2018}.

Then, the Fe and Cu diffusion coefficients are obtained by directly simulating the tracer diffusion experiments with MC simulations \citep{Mishin1997,Murch1981}. We compute the mean square displacement of the tracers ($<r^{2}>$) and the physical time at each temperature (eq. \ref{selfdiff2}).  

The physical time $t$ is re-scaled in order to consider the equilibrium vacancy concentration instead of the actual vacancy concentration of the simulation (the diffusion coefficient is multiplied by the factor $C_{v}/C_{MC}$). \citep{Soisson2007,Martinez2012}

During these MC simulations, at each $T$, we start performing $5\cdot 10^8$ spin Metropolis MC steps to reach the equilibrium magnetic state, then 600 spin steps are performed after each atomic MC step, consisting in a $1nn$ atom-vacancy exchange based on a time residence algorithm. 
For simplicity, we assume the typical time spent for one atom-vacancy exchange is sufficiently short, so that all the atomic spins remain frozen while going from the initial to the saddle-point state. However, we have determined that considering another assumption has a negligible effect on the results. Indeed, similar simulations were performed assuming the opposite, being that the spin-variation time is much shorter than the lifetime of both the initial and the saddle-point states, and very close migration barrier were obtained \citep{SM}. This test suggests that these properties are not sensitive to the detailed way of implementing the characteristic time of spin variations, and contribute to support the validity of our results. 

Via the same MC simulations, we also obtained the magnetic free energy of vacancy migration ($G^{m}_{mag}$) in the self-diffusion case, and the magnetic free energy barrier for Cu-vacancy ($G^{m,Cu}_{mag}$) and the distinct Fe-vacancy exchanges in the Cu diffusion case.

%RESULTS

%Magnetic free energies, enthalpies and entropies; Quantum statistics effects:

\begin{figure}[htbp]
  \centerline{\includegraphics[width=1\linewidth]{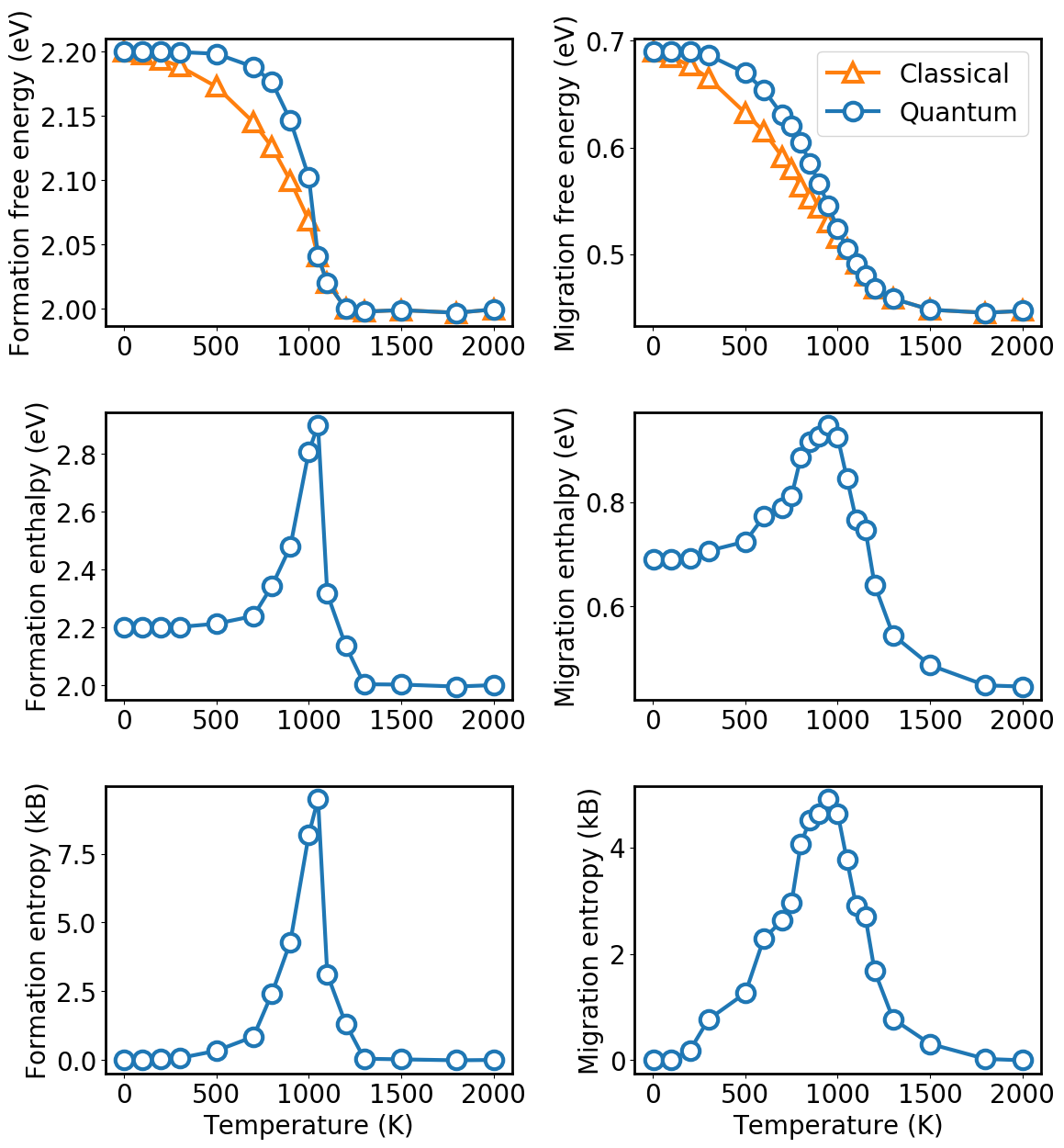}}
  \caption{Left panel: Magnetic free energy, enthalpy and entropy of formation. Right panel: Magnetic free energy, enthalpy and entropy of migration.}
  \label{figure_1}
\end{figure}

Fig. \ref{figure_1} shows the obtained magnetic free energy for vacancy formation and migration in pure iron, comparing results applying the Bose-Einstein distribution and the Boltzmann statistics for spin-MC at low temperatures. The difference between the two curves is significant, the slope approaching to zero near $T = 0$ only with the former approach. Therefore the quantum statistics is necessary to obtain a correct low temperature behavior of the vacancy formation and migration entropies (Fig. \ref{figure_1}), in agreement with previous spin-lattice dynamics data \citep{Wen2016}. 

The vacancy formation and migration energies at perfect FM and PM states are listed in Table \ref{table}, together with the resulting activation energies. At the low temperature limit, we reproduce closely the DFT energies (see \ref{table}). The asymptotic PM energies are obtained at 2000 K. These values for pure Fe are in good agreement with previous DFT and available experiments values \citep{DeSchepper1983,Ding2014,SM} 

In the FM state, the values for iron and for Cu diffusion are clearly different due to the Cu-vacancy attraction (0.24 eV and 0.17 eV for respectively $1nn$ and $2nn$ distances). However, such differences become significantly smaller in the PM state, indicating the dominance of the magnetic disorder over the chemical effect in the very dilute Fe-Cu system.

\begin{table} [htbp]
\begin{center}
\begin{tabular}{ccccc}
  \hline
  \hline
   & Fe$_{FM}$ & Fe$_{PM}$ & Cu$_{FM}$ & Cu$_{PM}$ \\
  \hline
  $H^{f}$ & 2.20 & 1.99 & 2.04 & 1.99 \\
  $H^{m}$ & 0.69 & 0.43 & 0.55 & 0.39 \\
  $Q$ & 2.92 & 2.46 & 2.66 & 2.44 \\
  $Q$ (exp.) & 2.63-3.10$^{a}$ & 2.48-2.92$^{a}$ & 2.53$^{b}$ & 2.43$^{b}$ \\
  \hline
  \hline
\end{tabular}
\caption{\label{table}
Values (in eV) of formation, migration and activation free energies for Fe and Cu diffusion in bcc iron at the ferromagnetic and paramagnetic limits. The error-bars due to the fitting are estimated to 0.05 eV. Diffusion pre-factors for Fe and Cu diffusion are respectively $1.8\cdot 10^{-4} (\pm9\cdot 10^{-5})$ m$^2$.s$^{-1}$ and $6.7\cdot 10^{-5} (\pm3\cdot 10^{-6})$ m$^2$.s$^{-1}$. (a) are obtained from Refs. \onlinecite{Graham1963,Walter1969,Buffington1961,James1966, Hettich1977,Geise1987,Iijima1988,Luebbehusen1988,DeSchepper1983}. (b) are obtained by an arrhenius fitting of the diffusion data from Ref. \onlinecite{Toyama2014}.}
\end{center}
\end{table}

%Comparison with experiments:

\begin{figure}[htbp]
  \centerline{\includegraphics[width=1\linewidth]{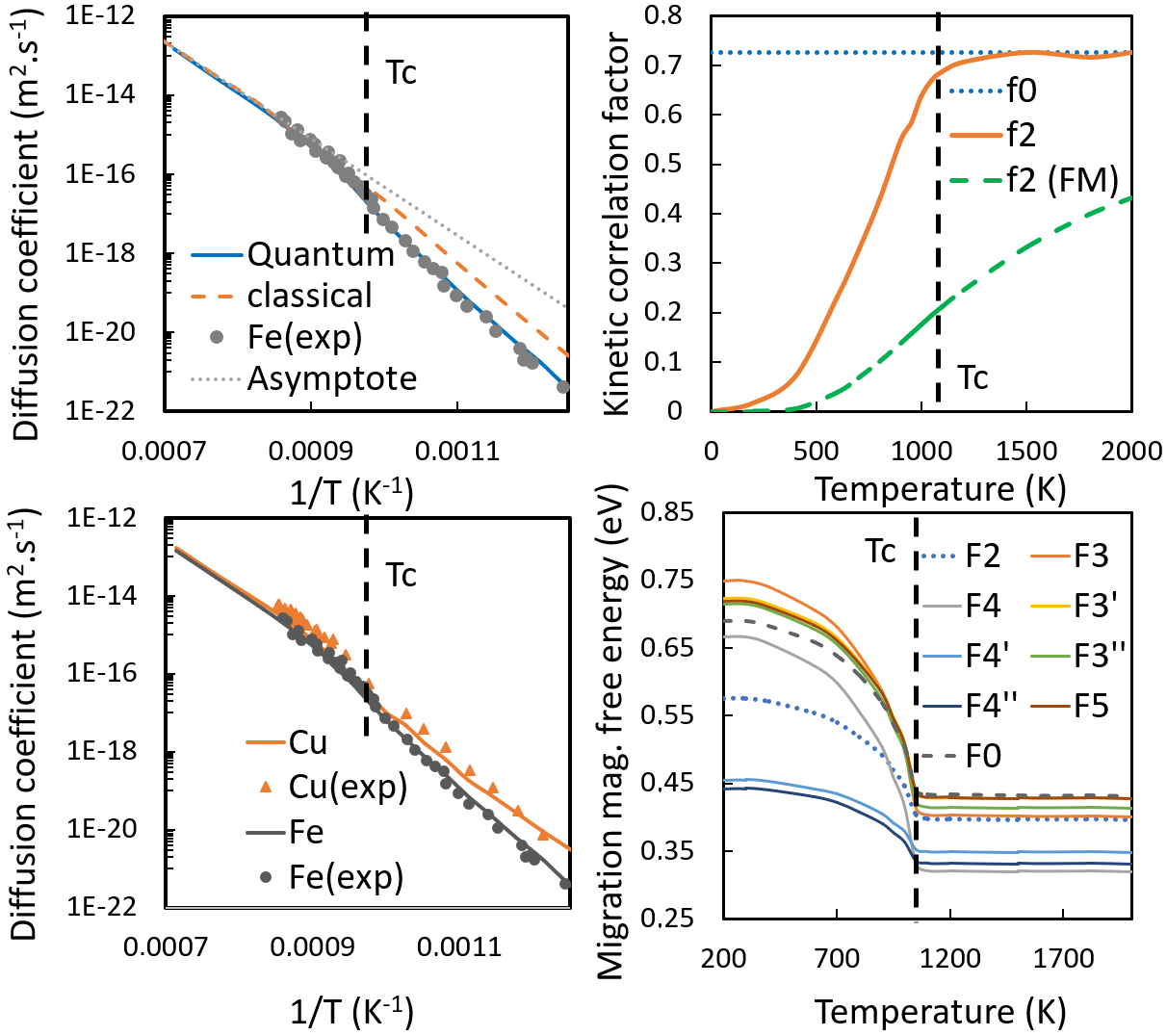}}
  \caption{Upper-left panel: Self-diffusion coefficients of Fe versus T, with and without quantum effects and comparison with experimental data obtained from \citep{Graham1963,Walter1969,Buffington1961,James1966, Hettich1977,Geise1987,Iijima1988,Luebbehusen1988}. Bottom left panel: Diffusion coefficients of Fe and Cu, compared with experimental results, obtained from \citep{Graham1963,Walter1969,Buffington1961,James1966, Hettich1977,Geise1987,Iijima1988,Luebbehusen1988} for Fe and \citep{Iijima1988,Toyama2014,Salje1977} for Cu. Upper-right panel: Kinetic correlation factors of Fe self-diffusion (f0) and Cu diffusion (f2). The f2 kinetic correlation with the frozen FM state (see text) is also displayed. Bottom right panel: Migration magnetic free energies of various jumps.}
  \label{figure_2}
\end{figure}

As shown in Fig. \ref{figure_2}, the present approach predicts the self- and Cu-tracer diffusion coefficients as functions of temperature, in excellent agreement with experimental studies \citep{Graham1963,Walter1969,Buffington1961,James1966, Hettich1977,Geise1987,Iijima1988,Luebbehusen1988, Speich1966, Rothman1968, Lazarev1970, Salje1977}. Especially, the sudden deviation from Arrhenius law near the Curie temperature is consistently predicted without any additional assumption. The change of slope (activation energy, $Q$) between the ferromagnetic and the paramagnetic regimes is also successfully predicted.

The self-diffusion coefficients obtained with purely classical statistics are also shown for a comparison. It reveals that using the Boltzmann distribution at low T in the spin-MC simulations significantly underestimates the local acceleration of diffusion around the Curie point.

Fig. \ref{figure_2} also shows that the kinetic correlation factor $f_2$ for Cu diffusion increases with temperature up to an asymptotic limit of 0.73, which is the $f_0$ value in pure bcc iron. 
To clarify the role of magnetic disorder on the $f_2$, we have performed similar MC simulations for Cu diffusion, but imposing a perfect FM order for all atomic-MC temperatures. The results show that the kinetic correlation factor of Cu diffusion increases more slowly when magnetic disorder is absent, This comparison together with the smaller difference between different barriers at the PM than at the FM state (Table \ref{table} and Fig. \ref{figure_2}) suggest the dominance of the magnetic disorder over the chemical interaction effect in this very dilute Fe-Cu system. Indeed, the magnetic free energy of binding between a vacancy and a Cu atom at 1$nn$ decays from 0.24 eV in the FM state (consistently with our DFT results) to 0.09 eV in the PM state.

%Validation of the Ruch model

\begin{figure}[htbp]
  \centerline{\includegraphics[width=1\linewidth]{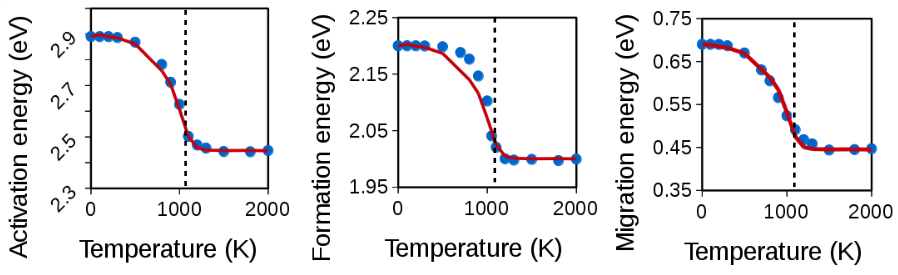}}
  \caption{Activation, formation and migration free energies compared with the Ruch model.}
  \label{figure_3}
\end{figure}

It is worth noting that if applying the Ruch model using the currently obtained $Q_{PM}$ and $Q_{FM}$ and $S(T)$, we obtain very similar diffusion coefficients.
For a closer comparison between results from our method and by using the Ruch model for an interpolation, Fig. \ref{figure_3} shows the respective data for the activation free energy, and the magnetic free energy of vacancy formation and migration.
As can be seen, both methods give very close values, especially above the Curie temperature. The largest differences occur between the formation free energies at $T < T_{C}$. In any case, the discrepancies are smaller than 10 \%. This good agreement between our approach and the Ruch model shows that the effects of magnetic short range order on vacancy properties are rather limited. In the case of Cu solute diffusion, the Ruch model also provides very close results. \citep{SM}

These comparisons suggest that the simple Ruch model allows a very good description of the temperature evolution of activation, formation and migration free energies in a ferromagnetic system. However, it should be noted that the Ruch interpolation requires an accurate knowledge of the asymptotic (FM and PM) energetic values, and the temperature evolution of the magnetization. They are generally not obvious to obtain both experimentally or from \textit{ab initio} calculations, especially for alloys beyond the dilute limit. Also, if estimating the PM energies for  highly itinerant magnetic systems, such as fcc Ni or bcc Cr via DFT, local-magnetism constraints should be applied, which can be extremely time consuming \cite{Abrikosov2016,Gambino2018}.  

%Advantages and approximations of this approach:

In summary, we propose an approach to efficiently predict atomic diffusion properties in iron, by performing on-lattice Monte Carlo simulations using effective interaction models parameterized on DFT data. These EIMs contain explicitly both chemical and magnetic variables, and the presence of a vacancy.

This approach naturally accounts for the interplay between magnetic and chemical degrees of freedom.

It is shown to successfully predict the temperature evolution of vacancy formation and migration magnetic free energies, and the tracer diffusion coefficients of Fe and Cu in bcc iron, across the Curie temperature. This approach is also ready to address the diffusion as a function of solute concentrations, and in the presence of non-equilibrium vacancies. The same approach is fully transferable to other magnetic metal systems.

At variance with the DFT-Ruch method, the current approach predicts properties for all temperatures regardless of the magnetic state. The crucial issue is to accurately parameterize the EIMs. 
On the other hand, it allows to reach to a calculation time of several orders of magnitudes longer than the spin-lattice dynamics. Therefore, it is also promising to address more complex kinetic processes than the atomic diffusion, such as the ordering, precipitation or segregation.

\acknowledgments

This work was partly supported by the French-German ANR-DFG MAGIKID project. Ab-initio calculations were performed using DARI-GENCI resources under the A0070906020 project and the CINECA-MARCONI supercomputer within the SISTEEL project.

\bibliographystyle{apsrev}
\bibliography{Schneider_resub2}

\clearpage

\end{document}